\documentstyle[epsfig,12pt]{article}
\newlength{\dinwidth}
\newlength{\dinmargin}
\newcommand{\ba}{\begin{array}}
\newcommand{\ea}{\end{array}}
\newcommand{\be}{\begin{equation}}
\newcommand{\ee}{\end{equation}}
\newcommand{\bea}{\begin{eqnarray}}
\newcommand{\eea}{\end{eqnarray}}

\def\a{\alpha}

\def\p{\pi}

\def\l{\lambda}

\def\G{\Gamma}

\def\L{\Lambda}
\def\to{\rightarrow}
\begin{document}
\thispagestyle{empty}
\addtocounter{page}{-1}
\begin{flushright}
SLAC-PUB-7245\\
SNUTP 95-053\\
{\tt hep-ph/9608247}\\
August 1996
\end{flushright}
\vspace*{1.3cm}
\centerline{\Large\bf $V_{ub}$ from the Hadron Energy Spectrum}
\centerline {\Large\bf in}
\centerline{\Large\bf Inclusive Semileptonic B Decays
\footnote{Work supported in part by Schweizerischer
Nationalfonds and the Department of Energy Contract
DE-AC03-76SF00515, U.S.NSF-KOSEF Bilateral Grant,
KOSEF Purpose-Oriented Research Grant 94-1400-04-01-3 and SRC-Program,
KRF International Coorporation Grant and Nondirected Research Grant
81500-1341, Ministry of Education BSRI Grant 95-2418, and 
Seoam Foundation Fellowship.}}
\vspace*{1.7cm}
\centerline{\large\bf Christoph Greub$^a$ and Soo-Jong Rey$^{a,b,c}$}
\vspace*{0.4cm}
\centerline{\large\it Stanford Linear Accelerator Center$^a$
\& Physics Department$^b$}
\centerline{\large\it Stanford University,
Stanford, California 94309, USA}
\vspace*{0.4cm}
\centerline{\large\it
Physics Department \& Center for Theoretical Physics$^c$}
\centerline{\large\it Seoul National University, Seoul 151-742 Korea}
\vspace*{2.1cm}
\centerline{\Large\bf Abstract}
A measurement of the hadron energy spectrum in inclusive
semileptonic $B$ decays is proposed as a
viable method for extracting $|V_{ub}|$.
Compared to the traditional 
energy spectrum
of the charged lepton, the hadron energy spectrum exhibits kinematical
advantages
such as a wider energy window and a larger signal branching fraction.
It is emphasized that the hadron energy spectrum
method is most suited for symmetric
$B$ factories, such as CLEO.
The hadron energy distribution is calculated in the approach of the
Altarelli et al. model and of the heavy-quark effective field theory.
In both methods,
perturbative QCD corrections, the Fermi motion of the
$b$-quark in the $B$-meson, and the recoil momentum of the $B$-meson
(stemming from the $\Upsilon(4S)$ resonance) are taken into account.
We have found excellent agreement between the spectra
calculated in both
methods, especially in the relevant kinematical region below the
charmed meson threshold. The theoretical error to
$|V_{ub}|$, which is dominated by the uncertainty of the $b$-quark mass,
is estimated to be at the $\pm 12 \%$ level.

\vspace*{1.1cm}
\centerline{Submitted to Physical Review D}

\newpage

\section{Introduction}
A precise determination of the Cabibbo-Kobayashi-Maskawa
matrix element
$V_{ub}$ is an important step for constraining the unitarity triangle.
Therefore, it
poses a challenge for both theory and experiment~\cite{alilecture}.
Traditionally, $V_{ub}$
has been extracted from the energy spectrum of the
charged lepton in inclusive
semileptonic $B$ decays  
$B \rightarrow X_u \ell  \nu$
above charm threshold, i.e.,
for lepton energies $E_\ell$ above $(m_B^2 - m_D^2)/(2m_B)
\approx 2.3$ GeV~\cite{cleoalt,argusvub}~\footnote{
We will discuss the impact on the determination of $|V_{ub}|$
from the recent CLEO measurements of the corresponding
exclusive channels
$B \to (\pi,\rho,\omega) \, \ell \nu$
\cite{CLEOexclalt,CLEOexclneu} later.}.
As the kinematical endpoint for semileptonic $B$ decays is at
$E_\ell=(m_B^2 - m_\pi^2)/(2 m_B) \approx 2.6$ GeV,
the pure $b \rightarrow u \ell^- {\overline \nu}$ transition
extends over a
relatively narrow window of about 300 MeV, accepting only
a fraction of approximately 10 \% of the total sample of charmless
$B$ decays. Theoretically, the lepton energy spectrum may be calculated
from  first--principle QCD. Using the
tools of the operator product expansion (OPE) and
heavy-quark effective field theory~\cite{hqetreview},
one can construct a systematic expansion of
the lepton spectrum in powers of $\L/m_b$, where $\L$
is a typical low-energy scale of QCD
\cite{hqetope}.
However, the relevant
lepton energy window from which $|V_{ub}|$
can be extracted lies to a large extent in
the so-called endpoint region
which extends from the parton model maximum
at $E_\ell=m_b/2$ up to the hadronic maximum at
$(m_B^2 - m_\pi^2)/(2m_B)$. It is therefore described by genuinely
non-perturbative contributions.
The difficulty arises from the fact~\cite{powercorr}
that, close to the partonic endpoint,
the expansion parameter is no longer $\L/m_b$, but $\L/(2m_b-E_\ell)$.
Thus, the theoretical prediction becomes
singular when the lepton energy approaches the parton model endpoint;
formally, these singularities manifest themselves in delta functions
and derivatives of delta functions concentrated at the partonic
lepton energy endpoint $m_b/2$.  In addition, this region
is plagued  by large perturbative Sudakov-like double
logarithms \cite{Sudakov,accmm,CCM} as well as by small instanton 
effects~\cite{chayrey}.
Therefore, in a large part of the lepton energy region in which 
the extraction of $V_{ub}$ is kinematically possible,
a full resummation of Sudakov-like double logarithms~\cite{accmm} and
power corrections~\cite{hqetresum}  becomes necessary.
In view of the theoretical and experimental difficulties
just mentioned, it is desirable to look for other  methods.

As an alternative,
we propose to extract $V_{ub}$ from the
hadron energy spectrum in the inclusive charmless
semileptonic $B$ decays $B \to X_u \ell \nu$.
Simple kinematical considerations support why this proposal
is viable. As the charmed final state
threshold is at the $D$-meson mass,
the maximal hadron energy window for
charmless semileptonic $B$ decays is given by the range
$m_\pi \le  E_{had}  \le m_D$; this window is much wider
than the corresponding kinematical
window for the lepton energy distribution
discussed above, and as we will see later, a much larger fraction
of the $B \to X_u \ell \nu$ events becomes  accessible,
 leading to improved statistics.
Of course, the theoretical problems addressed
in the discussion of the lepton endpoint spectrum  are also
present in principle in the hadron
energy spectrum~\cite{bouzas,falketal},
but to a much lesser extent in the kinematical region relevant
for the extraction of $V_{ub}$.
Indeed, up to perturbative QCD corrections, the region around
$E_{had}=m_b/2$ is also fully dominated by non-perturbative effects,
quite in analogy to the lepton energy endpoint region.
Fortunately, this region is well above the
charmed hadron final state threshold lying
outside the region we are interested in.
At the lower end of the hadron energy, $E_{had} \approx 0$,
the hadronic mass ranges over a narrow window
$0 \le m_{had}^2 \le E_{had}^2$, hence, the OPE breaks down again.
This region can be avoided by applying a lower cut to the
hadron energy; we choose the value 1 GeV.
This relatively high lower--cut has the advantage that a wide
range of invariant hadronic masses contributes
to the hadron energy spectrum; quark-hadron duality, which we
implicitly assume in our treatment for the inclusive
$B \to X_u \ell \nu$ decay,
is then expected to work well. Even after this cut at 1 GeV
is made, we are still left with an ample hadron energy window
ranging from 1 GeV to $m_D$.
These features in principle make investigations with hadron
energy spectra more reliable than with lepton energy spectra.

Experimentally, the hadron energy spectrum in semileptonic $B$ decays
may be measured schematically as follows:
Working at the $\Upsilon(4S)$ resonance,
which decays into $B \bar B$, one requires one of the $B$-mesons to
decay semileptonically and the other one hadronically. In the case of
a symmetric $B$-factory, like CLEO, the energy of the hadrons stemming
from the semileptonically decaying $B$-meson can be obtained by
measuring the  total  energy of all the hadrons in the final state
and then subtracting $m_{\Upsilon(4S)}/2$.
In case of asymmetric $B$ factories, the hadron
energy spectrum is harder to measure. One way is
to reconstruct in a first step the whole $\Upsilon(4S)$ decay
in its rest frame and then perform the analysis just described
for the symmetric case.
To perform the corresponding boost,
one has to measure precisely
both the energy and the momentum
of each final state hadron, which requires accurate particle
identification.

The remainder of this paper is organized as follows. In section 2
the calculation of the hadron energy spectrum is presented.
We utilize two methods to account for the bound-state effects.
The first one, discussed in section 2.1, uses the Altarelli et al.
(ACCMM) model~\cite{accmm} of the $B$-meson and
the second one, presented in 2.2, is based on the heavy-quark effective
field theory (HQET)~\cite{powercorr}.
In both treatments, the perturbative
$O(\a_s)$ corrections are taken into account.
In section 3 the extraction of $|V_{ub}|$ and its theoretical
errors are discussed.
Finally, in section 4 we give a brief
summary together with some comments on the experimental possibilites.

\section{Hadron Energy Spectrum}
In this section we present the calculation of the hadron energy spectrum
which is observed in the rest frame of the $\Upsilon(4S)$ resonance.
This frame coincides with the laboratory frame in the case of
a symmetric $B$-factory.
We take into account perturbative QCD corrections to $O(\alpha_s)$,
but resum Sudakov-like double logarithms.
Bound-state effects are evaluated within the
ACCMM model~\cite{accmm} and the HQET~\cite{powercorr} approaches.
We anticipate that both methods will lead to similar results
which, for the present case of a heavy to light transition, depend
effectively only on one parameter, viz. $b$-quark mass.
Theoretical accounts as to why the two distinct descriptions of
bound-state effects in heavy to light transitions should agree,
have been given in Ref.~\cite{accmmhqet}.

\subsection{Hadron Energy Spectrum in the ACCMM Approach}
In the ACCMM model the $\bar{B}$-meson consists of a $b$-quark and a spectator
antiquark $\bar q$, flying back-to-back in the $\bar{B}$-rest frame 
with 3-momentum vectors $\vec{p}$ and $-\vec{p}$, respectively; 
the momentum distribution $\Phi(p)$ of the spectator  is assumed to be 
of Gaussian form
\be
\label{wave}
\Phi(p) = \frac{4}{\sqrt{\pi} p_F^3} \, \exp
\left(- \frac{p^2}{p_F^2}\right) \quad ; \quad p =|\vec{p}| \quad ,
\ee
normalized according to
\be
\int_0^\infty \, dp \, p^2 \, \Phi(p) =1 \quad .
\ee
The basic feature of this model is the requirement that in the $\bar B$
rest frame the energies of the two constituents have to add up to the
$\bar B$-meson mass $m_B$. This is possible only when at least one of
the masses of the constituents is allowed to be 3-momentum dependent.
Usually, the mass of the spectator $m_{sp}$ is considered to be
fixed to a constituent quark mass
value, while the $b$-quark mass becomes
momentum dependent,
\be
\label{mbp}
m_b(p)^2 = m_B^2 + m_{sp}^2 - 2 m_B \,
\sqrt{p^2+m_{sp}^2} \quad .
\ee
We now consider the semileptonic decay
$ \bar B =(b,{\overline q}) \rightarrow u (+g) \ell \overline \nu +
\overline{q}$. The symbols
$p_b$, $p_u$, $p_g$, $p_\ell$, $p_{\overline \nu}$, $p_q$, $p_B$, and
$p_H$ denote the four-momenta of the $b$-quark,
the $u$-quark, the gluon,
the charged lepton, the anti-neutrino,
the spectator, the $B$-meson,
and the hadronic matter of the final state, respectively.
In addition, we define the four vector $q^\mu$
\be
q = p_\ell + p_{\overline {\nu}} \quad .
\ee
Due to the fact that $p_b^\mu + p_q^\mu = p_B^\mu$ in the ACCMM model,
the vector
$q^\mu$ is the same with or without the spectator four-momentum, viz.,
\be
q= p_B - p_H = p_b +p_q - p_u (- p_g) -p_q = p_b -p_u (-p_g) \quad .
\ee
For this reason, it is technically easier
to concentrate first on the double differential distribution
$d^2 \Gamma/dq^2dq^0$,
because the spectator does not appear explicitly.
To achieve this,
we first derive the double differential decay rate (including the
$O(\a_s)$ radiative corrections)
for $b \to u (+g) \ell {\overline \nu}$, where
the $b$-quark decays at rest.
For simplicity, the $u$-quark is assumed to be massless.
The kinematical variables $q^2$ and $q^0$ vary in the region
\bea
\label{rangeq2q0}
0 \le q^2 \le m_b^2 \quad , \nonumber \\
\left( q^2 \right)^{1/2} \le q^0 \le {m_b^2 + q^2 \over 2 m_b} \quad .
\eea
For each $q^2$, the tree-level process
(as well as the virtual gluon corrections)
are concentrated at the upper
endpoint of the $q^0$ range in Eq.(\ref{rangeq2q0}).
In the following it is more convenient to introduce
dimensionless variables $x$ and $y$ which vary
independently in the range
$[0,1]$:
\bea
q^2 = x^2 m_b^2, \nonumber \\
q^0 = [x + {1 \over 2} (1-x)^2 y] m_b.
\eea

With the above change of variables, we now proceed to calculate
the double differential distribution $d^2 \Gamma / dx dy$.
In terms of the new variables $(x,y)$, the tree-level contribution
and the virtual corrections are concentrated at $y=1$, while the
bremsstrahlung corrections give rise to a continuous distribution in
the whole $(x,y)$ domain. Individually, both the
virtual corrections and bremsstrahlung contributions
suffer from infrared and collinear divergences, which
occur at $y=1$ for a given value of $x$.
However, if the variable $y$ is integrated over a range
$s_0 \le y \le 1$ ($s_0 < 1$), these singularities cancel,
such that the quantity
\be
\label{pert}
{d \Gamma \over d x} (s_0) = \int_{s_0}^1 {d^2 \Gamma \over d x dy} d y
\nonumber \\
= \int_0^1 {d^2 \Gamma \over dx dy} dy - \int_0^{s_0} {d^2 \Gamma_{
brems} \over dx dy} dy
\ee
remains finite. The first term on the right hand side (RHS)
in Eq.(\ref{pert}) has been calculated  in Ref.~\cite{kuhn1};
the result is
\be
\label{contr1}
{d \Gamma \over d x} \equiv \int_0^1 {d^2 \Gamma \over dx dy} dy
= 4 x (1-x^2)^2 (1 + 2 x^2) \, [ 1 -{2 \alpha_s \over 3 \pi} G(x)]
\, \Gamma_0
\quad ,
\ee
where $\Gamma_0$ and the function $G(x)$
which containes the radiative corrections are given by
\bea
\label{gamma0}
\Gamma_0 &=& \frac{G_F^2 \, m_b^5 \, |V_{ub}|^2}{192 \pi^3}
\quad ,
 \\
G(x) &=&
\frac{[8x^2(1-x^2 -2x^4) \log x
+ 2(1-x^2)^2 (5 + 4 x^2) \log(1-x^2) -
(1-x^2)(5 + 9x^2 - 6 x^4) ]}{2 (1-x^2)^2 (1 + 2x^2)}
\nonumber \\
&& + \pi^2 + 2 \mbox{Li}(x^2) - 2 \mbox{Li}(1-x^2) \quad.
\eea
Here $\mbox{Li}(x)$ is the Spence function defined as
\be
\mbox{Li}(x) = - \int_0^x {dt \over t} \log(1-t).
\ee
Because the tree-level contribution and virtual
corrections are concentrated
at $y=1$, the second term on the RHS of Eq.~(\ref{pert})
contains the gluon bremsstrahlung contribution
only (as indicated by the notation).
As the endpoint region is cut off by $s_0 < 1$,
this term is finite;  consequently, the infrared and collinear
regularization is not necessary from the very beginning.
We have calculated this term; the result is
\bea
\label{contr2}
\int_0^{s_0} {d^2 \Gamma_{brems} \over d x d y} d y =
4 x (1 - x^2)^2 (1 + 2x^2) \, {2 \alpha_s \over 3 \pi}
\, \left( \log^2(1-s_0) + H(s_0) \right) \,
\Gamma_0 \quad ,
\eea
where
\bea
\label{h0}
H(s_0) = && \int_0^{s_0} dy \Big( {4 \over 1 - y}
\log{2 - y(1-x) + \kappa
\over 2}          \nonumber \\
&&- {(1-x)(3 + x + xy - y) \over (1+x)^2}
\Big[\log(1-y) - 2 \log{2 - y(1-x) + \kappa \over 2} \Big]
\nonumber \\
&&- { \kappa \over 2 (1 + x)^2 (1 + 2x^2) }
\Big[{7 (1+x) (1 + 2 x^2) \over 1 - y} + (1-x)(3 - 2 x^2) \Big]\Big).
\eea
The quantity $\kappa$ in Eq. (\ref{h0}) is defined as
$\kappa = \sqrt {y^2 (1-x)^2 + 4 xy}$.
Combining the two contributions, Eqs.(\ref{contr1}) and (\ref{contr2}),
we obtain
\be
{1 \over \Gamma_0} {d \Gamma \over d x} (s_0)
= 4 x (1-x^2)^2 (1 + 2x^2) \,
\left[ 1 - {2 \alpha_s \over 3 \pi} \log^2 (1 - s_0)
- {2 \alpha_s \over 3 \pi}
\left( G(x) + H(s_0) \right) \right] \, \Theta(1 - s_0).
\ee
The double logarithms arise from the soft and
collinear gluons and become
important as $s_0 \rightarrow 1$. Resumming these double logarithmic
terms to all orders, we get
\be
{1 \over \Gamma_0} {d \Gamma \over d x}(s_0)
= 4 x (1 - x^2)^2 (1 + 2 x^2) \,
\exp\left(-{2 \alpha_s \over 3 \pi} \log^2 (1 - s_0) \right) \,
\left[ 1 - {2 \alpha_s \over 3 \pi} \left( G(x)
+ H(s_0) \right) \right] \,
\Theta (1 - s_0).
\ee
This expression enables us to reproduce the Sudakov exponentiated
double-differential decay rate by differentiating with respect to $s_0$.
\bea
\label{doubleexpon}
{1 \over \Gamma_0} {d^2 \Gamma \over d x d y} &=&
 - {d \over d s_0} \Big( {1 \over \Gamma_0}
{d \Gamma \over d x}(s_0) \Big)_{s_0 = y}
=\nonumber \\
&& - 4 x (1 - x^2)^2 (1 + 2 x^2) \,
\exp\Big( - {2 \alpha_s \over 3 \pi} \log^2 (1 - y) \Big)
\nonumber \\
&\times & \left\{
{4 \alpha_s \over 3 \pi} {\log(1-y) \over (1-y)}
\Big[ 1 - {2 \alpha_s \over 3 \pi} \big( G(x) + H(y) \big) \Big]
-{2 \alpha_s \over 3 \pi} {d H \over d y}(y) \right\}
.
\label{doublediff}
\eea

To get the parton level hadron energy spectrum for a $b$ quark
decaying at rest, we first re-express
Eq.(\ref{doubleexpon}) in terms of the variables $(q^2,q^0)$ and then
integrate over $q^2$; this leads to the
distribution $d\Gamma/dq^0$. As the
hadronic energy $E_{had}$ is related to $q^0$ by $E_{had}=m_b - q^0$,
the spectrum $d\Gamma/dE_{had}$ for a $b$--quark decaying at rest
is readily obtained.
To get rid of the $b$--quark mass dependence $m_b^5$ in the decay width
(as seen e.g. in Eq.(\ref{gamma0})) and errors thereof,
we present in the following the
differential branching ratio $dBR/dE_{had}$
that is obtained by dividing $d\Gamma/dE_{had}$
by the theoretical semileptonic $b$--quark decay width $\Gamma_{sl}$
and multiplying this ratio by the experimentally measured semileptonic
branching ratio $BR_{sl}=(10.4 \pm 0.4) \%$ \cite{Gibbons}:
\be
\label{brdiff}
\frac{dBR(B \to X_u \ell \nu)}{dE_{had}} = \left(
\frac{1}{\Gamma_{sl}} \frac{d\Gamma}{dE_{had}} \right) \, BR_{sl} .
\ee
The semileptonic decay width,
neglecting the small $B \to X_u \ell  \nu$
contribution, reads
\be
\label{semileptonic}
\G_{sl} = \frac{G_F^2 \, m_b^5 \, |V_{cb}|^2}{192 \p^3} \,
g(m_c/m_b) \, \left( 1 -
\frac{2 \a_s(m_b)}{3 \p} f(m_c/m_b) \right) ,
\ee
where the phase space function $g(u)$ is defined as
\be
\label{gu}
g(u) = 1 - 8 u^2 + 8 u^6 - u^8 - 24 u^4 \log u \quad ,
\ee
and the radiative correction function in an approximate analytic form
is given by~\cite{CCM}:
\be
\label{fu}
f(u) = \left( \p^2 - \frac{31}{4} \right) \, (1-u)^2 + \frac{3}{2}
.
\ee
The result (based on the the
Sudakov improved differential branching ratio
as just derived) is shown by the dash-dotted line in Fig. 1 (using
$m_b=4.85$ GeV and $m_c=1.45$ GeV). To emphasize the effects of
the Sudakov resummation, we have also plotted
the result obtained in pure $O(\a_s)$ perturbation theory \cite{kuhn2},
i.e., without
exponentiation (short-dashed line).
This curve exhibits a (integrable) double logarithmic  divergence
when $E_{had}=m_b/2$ is approached from above.
The effects of the exponentiation
are therefore most prominent in the region
around $m_b/2$ as illustrated in Fig.~1, because
Sudakov-exponentiation suppresses the 
singularity just mentioned.
We stress, however,
that the effect of exponentiation is
negligibly small in the whole region
below the
charmed hadron threshold we are mostly interested in.
This should be contrasted with the case of
the  energy spectrum of the charged lepton near
the kinematical endpoint where these effect are most pronounced.

Another interesting aspect concerning
the radiative corrections to the hadron
energy spectrum is that the kinematical boundary of the hadron energy
depends on whether the final state contains bremsstrahlung gluons
or not.
In the absence of bremsstrahlung gluons,
the kinematical endpoint is at $m_b/2=2.425$ GeV as can be seen
form the long-dahed curve representing
the result without QCD corrections.
Fig.~1 clearly illustrates that, above $E_{had} = 2.425$ GeV, a
significant tail due to gluon bremsstrahlung, hence of
order $\alpha_s(m_b)$ in strength, is present in the spectrum.
\begin{figure}[htb]
\vspace{0.10in}
\centerline{
\epsfig{file=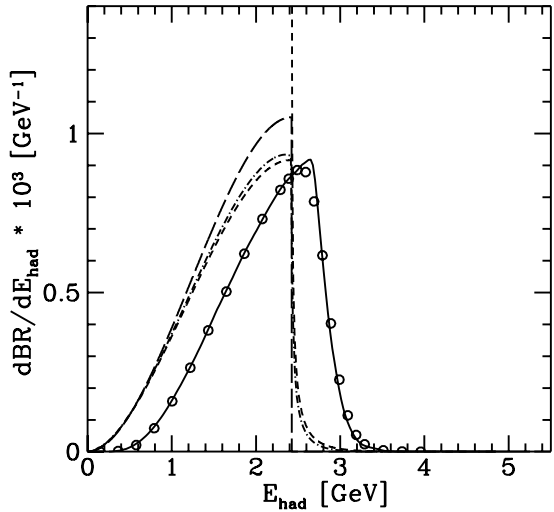,height=3in,angle=0,clip=}
}
\vspace{0.08in}
\caption[]{The long-dashed line shows the hadron energy spectrum for a 
$b$-quark decaying at rest without taking into account any QCD corrections.
The short-dashed line is the corresponding spectrum including
$O(\a_s)$ virtual and bremsstrahlung corrections. The result after
exponentiating the Sudakov double logarithms
(discussed in the present section) is shown by the dash-dotted line.
The solid line shows
the hadron energy spectrum for a $B$-meson decaying at rest.
The bound state effects are calculated with the ACCMM model
with $p_F=344$ MeV and $m_{sp}=150$ MeV
(corresponding to $\langle m_b \rangle =4.85$ GeV).
The open circles show the hadron energy spectrum due to a $B$-meson
decaying at flight with a momentum $|p_B|=330$ MeV.
\label{fig:1}}
\end{figure}

We now turn to implement the bound--state effects using the ACCMM model.
We start from the double differential distribution
$d \Gamma/dq^2dq^0$ in Eq.~(\ref{doubleexpon}). As the $b$-quark moves
with momentum $\vec{p}$ inside the $B$-meson,
we first replace the mass $m_b$ by $m_b(p)$ as given in Eq.~(\ref{mbp}).
We then Lorentz boost the double
differential distribution and get the spectrum for a $b$-quark
decaying at flight (momentum $\vec{p}$).
Finally we convolute the spectrum with the ACCMM distribution
function given in Eq.~(\ref{wave}). This  leads to the
ACCMM averaged quantity 
$d \Gamma/dq^2dq^0$ for a $B$-meson decaying at rest.
Again, this distribution is straightforwardly
converted to $d\Gamma/dE_{had}$.
To get the differential branching ratio, we replace
the $b$--quark mass $m_b$ in the expression
for the semileptonic decay width  in Eq. (\ref{semileptonic})
by the ACCMM average value $\langle m_b \rangle$ derived by applying
the ACCMM convolution to the total semileptonic decay rate.
This ACCMM averaged $b$-quark mass is given by
\be
\label{mbeff}
\langle m_b^5 \rangle = \int dp \, p^2 \,
\big(m_b(p)\big)^5 \, \Phi(p) \quad . \ee
The $m_b - m_c$ mass difference is reliably calculated to be 3.40 GeV
by HQET~\cite{hqetmass}. Therefore, we set
$m_c= \langle m_b \rangle - \, 3.40 \, \mbox{GeV}$ in the $\Gamma_{sl}$.
The result is shown in Fig. 1 by the solid line; here we have used
$p_F=344$ MeV and  $m_{sp}=150$ MeV,
which yields $\langle m_b \rangle =4.85$ GeV
according to Eq. (\ref{mbeff}).
As shown in Fig. 1, the main effect of the Fermi motion
of the $b$-quark inside $B$-meson is to
shift the perturbative hadron energy spectrum
uniformly to higher energies by about
300 MeV. We will elaborate this point in more
detail when discussing the HQET
approach to the bound-state effects, where the same result 
can be understood in a more transparent way.

There is one more source of Doppler shift to the hadron energy spectrum.
At a symmetric $B$-factory the $B$-meson is produced from the decay of the
$\Upsilon(4S)$ resonance. Due to the released binding energy
the $B$-mesons are produced with a momentum
$|\vec{p}_B| \approx 330$ MeV in the $\Upsilon(4S)$ rest frame.
We have also worked out this effect to the spectrum; the corresponding
result is shown in Fig. 1 by open circles.
As this effect is very small, we will not consider it any more
in foregoing discussions.

Although the ACCMM model has two input parameters, viz.
$p_F$ and $m_{sp}$,
it turns out that the hadron energy spectrum is sensitive only to
one parameter, namely, the average $b$--quark
mass $\langle m_b \rangle$,
which is a function of $p_F$ and $m_{sp}$
(see Eq.~(\ref{mbeff}))~\footnote{
It has been pointed out that transitions from heavy quarks  
to light  quarks with masses $m_q \le (\bar{\Lambda} m_b)^{1/2}$ always
give rise effectively to a one-parameter dependence~\cite{accmmhqet}.}.
To illustrate this point, we have chosen two different pairs of
$(p_F,m_{sp})$ values, which both correspond
to the same value of $\langle
m_b \rangle$
 (=4.85 GeV in the present case).
The correspondig result plotted in Fig.~2 indicates little dependence 
on $p_F$ and $m_{sp}$, once $\langle m_b \rangle$ is hold fixed.
In Fig.~3, we have varied  $\langle m_b \rangle$ over the
range indicated in the plot. We have found that the hadron energy
spectrum depends on $\langle m_b \rangle$ rather strongly,
especially in the
region below the charmed hadron threshold.

\begin{figure}[htb]
\vspace{0.10in}
\centerline{
\epsfig{file=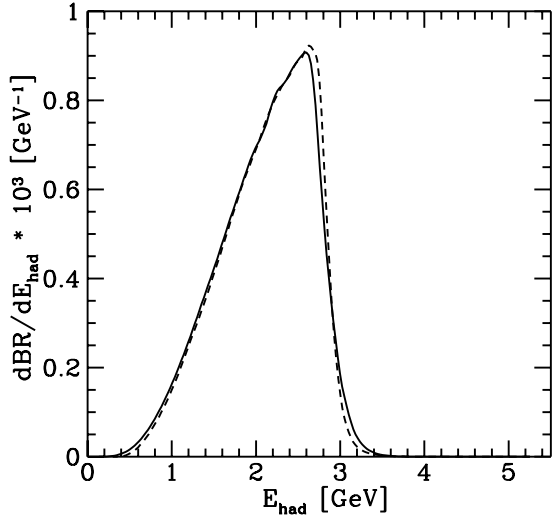,height=3in,angle=0,clip=}
}
\vspace{0.08in}
\caption[]{The hadron energy spectrum
(based on the ACCMM model) is shown for
two different pairs of model paramters $(p_F,m_{sp})$.
The solid  line corresponds to
$(374 \ \mbox{MeV}, 0)$ while the dashed line represents the result for
$(252 \ \mbox{MeV}, 300 \, \mbox{MeV})$.
Both set correspond to the same value of the $b$-quark mass $\langle m_b
\rangle =4.85$ GeV.
\label{fig:2}}
\end{figure}
\begin{figure}[htb]
\vspace{0.10in}
\centerline{
\epsfig{file=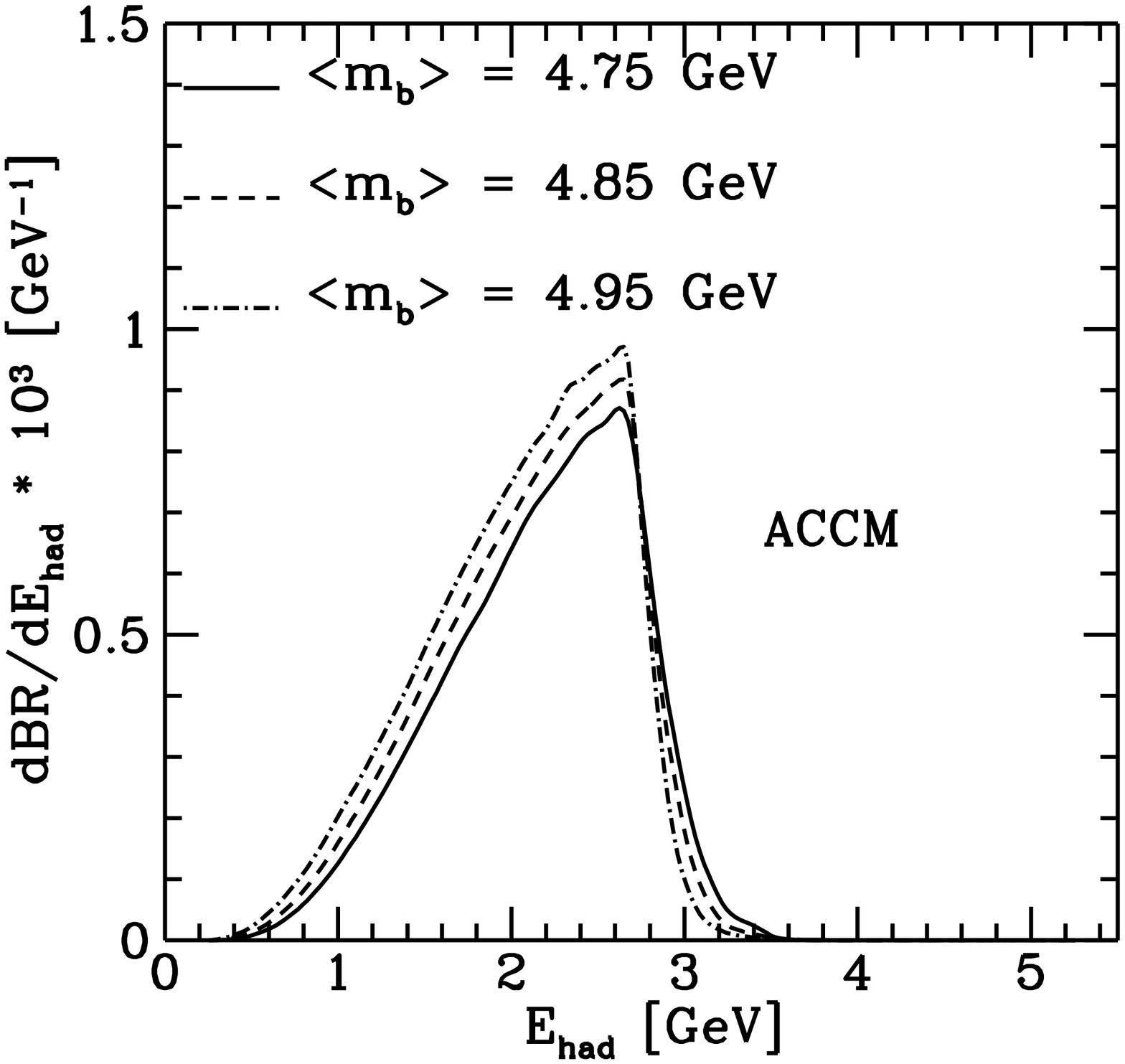,height=3in,angle=0,clip=}
}
\vspace{0.08in}
\caption[]{The hadron energy spectrum
based on the ACCMM approach is shown
for different values of the $b$-quark mass $\langle m_b \rangle$.
\label{fig:3}}
\end{figure}
\subsection{Hadron Energy Spectrum in the HQET Approach}
Next we calculate the hadron energy spectrum using the
HQET approach. Since the derivation is
discussed in detail in~\cite{falketal},
we only repeat the results relevant for our discussions.
In the following we take into account the leading corrections
in $\a_s$ and $1/m_b$, but we neglect
mixed higher--order corrections of the
order $\a_s \frac{1}{m_b}$. We assign the $b$ quark the same
four velocity $v^\mu$ as the heavy meson and define the dimensionless
parton level quantity $\hat{E}_0$ as
\be
\label{Ehut0}
\hat{E}_0 = \frac{v \cdot (p_b-q)}{m_b} = 1 - \frac{q \cdot v}{m_b} 
\quad .
\ee
The physical hadron energy $E_{had}$ is given as
\be
\label{Ehadron}
E_{had} = v \cdot (P_B-q) = m_B - q \cdot v  \quad .
\ee
The leading order equality between $m_B \hat{E}_0$ and $E_{had}$
gets modified by corrections linear in $1/m_b$
\be
\label{erelation}
E_{had} = \bar \L - \frac{\l_1+3\l_2}{2 m_B} +
\left(m_B - \bar \L + \frac{\l_1+3\l_2}{2 m_B} \right) \,
\hat{E}_0 + \cdots \quad ,
\ee
where we have used the relationship
\be
\label{mrelation}
m_B = m_b +\bar \L - \frac{\l_1+3\l_2}{2 m_b} + \cdots
\ee
between the $b$ quark and the $B$-meson masses.
The ellipses in Eqs.~(\ref{erelation}) and (\ref{mrelation})
stand for higher order corrections in the $1/m_b$ expansion.
We split the differential spectrum
$\frac{d\G}{d\hat{E}_0}$ into two parts:
\be
\label{split}
\frac{d\G}{d\hat{E}_0} =
\frac{d\G^{1/m_b^2}}{d\hat{E}_0} +
\frac{d\G^{\a_s}}{d\hat{E}_0} \quad .
\ee
The first term on the right hand side (RHS) of Eq.~(\ref{split})
consists of the $1/m_b^2$ corrections and
has been calculated in \cite{falketal}. For $m_u=0$ one gets
($0 \le\ \hat{E}_0 \le 1/2$)
\bea
\label{boundcorr}
\frac{1}{\G_0} \, \frac{d\G^{1/m_b^2}}{d\hat{E}_0}
&=& 16\, \hat{E}_0^2 \, [3-4\hat{E}_0]
\nonumber \\
&+&
 16 \, \left[ \frac{\l_1}{2m_B^2} \, \left(
-6 \hat{E}_0 + 12 \hat{E}_0^2 + \frac{20}{3} \, \hat{E}_0^3
\right) +
\frac{\l_2}{2m_B^2} \, \left( -3 - 6 \hat{E}_0 + 36 \hat{E}_0^2
+ 20 \hat{E}_0^3 \right) \, \right] \nonumber \\
&+& \left( \frac{5 \l_1}{3m_B^2} - \frac{\l_2}{m_B^2} \right) \,
\delta(\hat{E}_0 -1/2) + \frac{\l_1}{6m_B^2} \,
\delta '(\hat{E}_0 - 1/2) +O(\a_s;1/m_b^3) \quad ,
\eea
with
$\G_0$ given in Eq.~(\ref{gamma0}).
The second term on the RHS of Eq. (\ref{split}) contains
the perturbative $O(\a_s)$ corrections; it has been
calculated by Czarnecki et al. \cite{kuhn2}. 
For completeness, we list their result for a massless $u$-quark:
\be
\frac{1}{\G_0} \, \frac{d \Gamma^{\a_s}}{d \hat{E}_0} =
\frac{32 \a_s}{3 \pi} \, {\cal G}_1(\hat{E}_0) \quad ,
\ee
where the function ${\cal G}_1$ for $0 \le \hat{E}_0 \le 1/2$ reads
\bea
\label{calg1low}
{\cal G}_1(\hat{E}_0) &=& \hat{E}_0^2 \left\{\frac{1}{90}
 \left(16 \hat{E}_0^4 -
 84 \hat{E}_0^3 + 585 \hat{E}_0^2 - 1860 \hat{E}_0 +1215 \right)
\right. \nonumber \\
&& \left. + (8 \hat{E}_0-9) \, \log 2 \hat{E}_0 +
2(4\hat{E}_0 -3) \, \left[
\frac{\pi^2}{2} + \mbox{Li}(1-2\hat{E}_0) \right] \, \right\} \quad ,
\eea
while for $1/2 < \hat{E}_0 \le 1$
\bea
\label{calg1high}
{\cal G}_1(\hat{E}_0) &=& \frac{1}{180} \, (1-\hat{E}_0) \,
\left(32 \hat{E}_0^5 - 136 \hat{E}_0^4 +
 1034 \hat{E}_0^3 - 2946 \hat{E}_0^2 + 1899 \hat{E}_0 +312 \right)
 \nonumber \\
&& - \frac{1}{24} \, \log (2\hat{E}_0 -1 ) \,
\left(
64 \hat{E}_0^3 - 48 \hat{E}_0^2 - 24 \hat{E}_0 - 5
\right) \nonumber \\
&&
+ \hat{E}_0^2 \,(3-4\hat{E}_0) \, \left[ \frac{\pi^2}{3} -
4 \mbox{Li}(1/(2\hat{E}_0)) + \log^2(2 \hat{E}_0 -1 )
- 2 \log^2 (2 \hat{E}_0) \,
\right] \quad .
\eea
Using relation (\ref{erelation}) between $\hat{E}_0$\
and $E_{had}$, the corresponding hadron energy spectrum
$\frac{1}{\G_0}\frac{d\G}{dE_{had}}$ is readily obtained.
While the quantity
$\frac{1}{\G_0} \frac{d\G}{d\hat{E}_{0}}$
is free from corrections linear
in $1/m_b$, there are such corrections in the hadron energy spectrum
$\frac{1}{\G_0} \frac{d\G}{dE_{had}}$.
The differential branching ratio in the HQET approach is then
given by
\be
\label{hqetbr}
\frac{dBR( B \to X_u e  \nu)}{dE_{had}} =
\frac{|V_{ub}|^2}{|V_{cb}|^2} \,
\frac{\left( \frac{1}{\G_0}\frac{d\G}{dE_{had}} \right)} {g(m_c/m_b)} \, 
BR_{sl} \, \left[ 1 - \frac{2\a_s(m_b)}{3\pi} \, f(m_c/m_b) +
\frac{h(m_c/m_b)}{2m_b^2}
\right]^{-1} \quad ,
\ee
where the phase space function $g(u)$ and the
radiative correction function $f(u)$ are listed in Eqs.~(\ref{gu})
and (\ref{fu}), respectively. The function $h(u)$ contains
the $1/m_b^2$ corrections to the semileptonic decay
$B \to X_c \ell \nu$; it reads
\be
\label{hfun}
h(u) = \l_1 + \frac{\l_2}{g(u)} \,
\left[
-9+24u^2-72u^4+72u^6-15u^8-72u^4 \log u  \right] \quad .
\ee

To illustrate the qualitative features of
the non-perturbative corrections,
it is instructive to retain only the dominant terms which are linear
in $1/m_b$. The relation between $E_{had}$ and $\hat{E}_0$ (given in 
Eq.~(\ref{erelation})) then simplifies to 
$E_{had} = \bar \L + m_b \hat{E}_0 = \bar \L + E_0$, 
where $E_0$ is the hadron energy at the parton level.
Therefore, the dominant bound--state effect
is simply a uniform shift of the parton level hadron energy 
spectrum to higher energies by an amount of $\bar \L$; 
exactly the same feature was also noted  
when implementing the bound state effects using
the ACCMM approach (see Fig.~1).
{}From these considerations it is clear that
the most important parameter
is $\bar \L \approx m_B - m_b$, while the parameters $\l_1$
and $\l_2$ play a relatively minor role. In the numerical evaluations
we therefore have used the fixed values $\l_1=-0.5 \, \mbox{GeV}^2$ and
$\l_2=0.12 \, \mbox{GeV}^2$ and presented the results as a function 
of $m_b$ (or, equivalently, $\bar \L$) only.

In Fig.~4 we compare the ACCMM and the HQET approaches. As an example
we have chosen $m_b= \langle m_b \rangle =4.85$ GeV.
The plot illustrates that
the two methods agree with each other rather well, especially in the
energy range $\mbox{1 GeV}  \le  E_{had}  \le  m_D$.
We have checked that this agreement holds everywhere in the relevant
$b$ quark mass range $m_b=(4.8 \pm 0.2)$ GeV.
\begin{figure}[htb]
\vspace{0.10in}
\centerline{
\epsfig{file=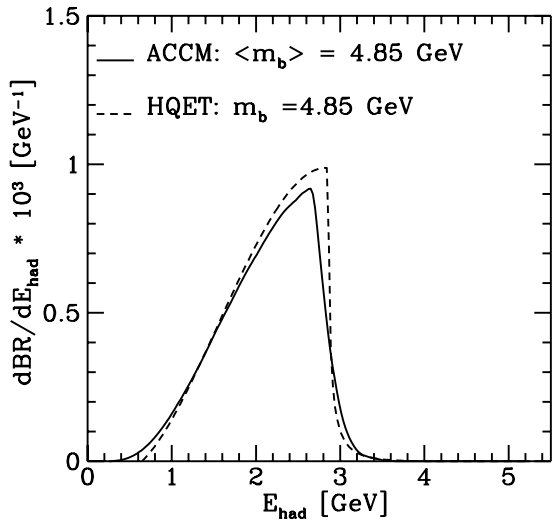,height=3in,angle=0,clip=}
}
\vspace{0.08in}
\caption[]{The hadron energy spectrum is shown for two different
descriptions of the bound state effects. The solid curve shows the
result obtained by using the ACCMM model while the dashed line
shows the HQET result.
\label{fig:4}}
\end{figure}
\section{Extraction of $V_{ub}$ from the Hadron Energy Spectrum}
As discussed in the Introduction,
theoretical predictions for the hadron energy spectrum
are expected to be rather reliable in the relevant kinematical region
below the $D$-meson mass, once a sufficiently high lower--cut in the 
hadron energy is made at the same time. 
Choosing the lower cut at 1 GeV, the invariant hadronic mass
ranges from $m_\pi$ up to $E_{had} \ge 1 \, \mbox{GeV}$ for all values
of the hadronic energy in the window
$1 \, \mbox{GeV} \le E_{had} \le m_D$.
In this case a large number of different
hadronic final states contributes
to the spectrum; hence, quark-hadron duality should hold quite well.
We therefore propose to measure experimentally the
kinematical branching ratio defined by
\be
\label{brkin}
BR^{kin} ( B \to X_u \ell \nu) =
\int_{1 \, GeV}^{m_D} \, \frac{dBR(B \to X_u \ell \nu)}{
dE_{had}} \, dE_{had} \quad .
\ee
A comparison with the corresponding theoretical quantity then allows
to extract the CKM ratio $|V_{ub}|/|V_{cb}|$. While leaving aside
the experimental question how accurately the kinematical branching
ratio can be measured now or in the future, we would like to point out
in the following that the theoretical uncertainties in this
observable are small enough to reduce substantially the present
theoretical error on $|V_{ub}|$.
As it turns out that the theoretical error of the kinematical
branching ratio is dominated by the uncertainty of the
effective $b$ quark mass (i.e., $m_b$ in the HQET approach and
$\langle m_b \rangle$ in the ACCMM approach),
we take into account only this effect in the following analysis
\footnote{The dependence of the kinematical branching ratio 
on the numerical value
of $\a_s$ is very small. In our numerical evaluations we have taken
$\a_s=0.205$.}. 
A reasonable range for the value of $m_b$ can be inferred from the 
measurement of the lepton energy spectrum in the inclusive decays
$B \to X_{c,u} \ell  \nu$. Fitting to the ACCMM model,
CLEO extracted $p_F$ to be $264 \pm 16$ MeV (using a value of
$m_{sp}=150$ MeV for the mass of the spectator quark)
\cite{cleoalt,Poling}. In a more recent
analysis using lepton tags \cite{cleotag}, they extracted the value
$p_F=347 \pm 68$ MeV. We thus choose to vary $p_F$ in the somewhat
larger range $200 \, \mbox{MeV} \le p_F \le 435 \, \mbox{MeV}$
which covers both measurements.   Using
Eq.~(\ref{mbeff}), this $p_F$-range translates into the
$\langle m_b \rangle$-range
\be
\label{mbrange}
4.75 \, \mbox{GeV} \le \langle m_b \rangle \le 5 \, \mbox{GeV}.
\ee
We note that this range is also compatible with the best value
($\langle m_b \rangle$=4.77 GeV)
fitted from the photon energy spectrum in $B \to X_s \gamma$
\cite{aligreub}.

So far, we have not discussed the question,
if the cascade decay $b \to c X \to s \ell \nu X$ , where
the symbol $X$ denotes light quarks, can fake a $b \to u \ell \nu$
transition. If the $c$-quark is off-shell, this process can
kinematically fake a $b \to u \ell \nu$ transition, but
in this case the process is higher order in the weak interaction,
hence
negligible. On the other hand, if the $c$ quark is on-shell,
the energy of the hadronic system $X$ is larger than $m_D$.
Therefore, this cascade process is not a backgound to the 
$b \to u \ell \nu$ transition; 
this is in contrast to the endpoint analysis of the lepton 
energy spectrum, where the cascade process has to be subtracted out.

To get an idea what fraction $R$ of the total semileptonic
$b \to u$ events will be captured in the energy window
$1 \, \mbox{GeV} \le  E_{had}  \le  m_D$, we
have used the present central
value for $|V_{ub}|/|V_{cb}|=0.08$ \cite{PDG96}.
The $m_b$ dependence of the kinematical branching ratio and of the
fraction $R$ are shown in Table 1 for both the ACCMM and the HQET
approaches.
\begin{table}[htb]
\label{tabelle}
\begin{center}
\begin{tabular}{| c || c | c || c | c | }
\hline
 $m_b$ (GeV) & $BR^{kin}_{ACCMM}*10^4$ & $R_{ACCMM}$ (\%)
                              & $BR^{kin}_{HQET}*10^4$
                              & $R_{HQET} (\%)$\\
\hline \hline
4.60  & $2.41$ & $23$ & $2.17$ & $21$ \\
4.65  & $2.58$ & $24$ & $2.40$ & $22$ \\
4.70  & $2.76$ & $25$ & $2.64$ & $24$ \\
4.75  & $2.94$ & $26$ & $2.89$ & $25$ \\
4.80  & $3.15$ & $27$ & $3.14$ & $27$ \\
4.85  & $3.37$ & $28$ & $3.41$ & $28$ \\
4.90  & $3.59$ & $29$ & $3.68$ & $29$ \\
4.95  & $3.83$ & $30$ & $3.95$ & $31$ \\
5.00  & $4.07$ & $31$ & $4.23$ & $32$ \\
\hline
\end{tabular}
\end{center}
\caption[]{The kinematical branching ratio $BR^{kin}$ defined
in Eq.~(\ref{brkin}) and the corresponding fraction $R$
of the semileptonic $b \to u$ events lying in the energy
window $1 \, \mbox{GeV}  \le  E_{had}  \le  m_D$ are given as a function
of $m_b$ for the ACCMM and the HQET approach.
$|V_{ub}|/|V_{cb}|=0.08$ is assumed.
\label{table1}}
\end{table}
Varying $m_b$ in the range specified in Eq. (\ref{mbrange}) 
and using the results
for $BR^{kin}_{HQET}$ in Table 1 (where the variation is
somewhat larger than in the ACCMM model), we obtain
\be
\label{predtheory}
BR^{kin} = \frac{|V_{ub}|^2}{|V_{cb}|^2} \, \times
(4.51 - 6.61) \, \times 10^{-2} \quad .
\ee
Denoting the measured kinematical branching ratio by $BR_{exp}^{kin}$,
one can extract $|V_{ub}|/|V_{cb}|$ to be
\be
\label{vubres}
\frac{|V_{ub}|}{|V_{cb}|} = \sqrt{BR^{kin}_{exp}} \, \cdot
\, (4.30 \pm 0.41) \quad .
\ee
This implies that the theoretical error of the ratio
$|V_{ub}|/|V_{cb}|$ is approximately $\pm 10\%$.
Taking the somewhat larger range $m_b=(4.80 \pm 0.15)$ GeV
adopted in~\cite{alilecture,bagan},
we get from Table 1
\be
\label{vubres1}
\frac{|V_{ub}|}{|V_{cb}|} = \sqrt{BR^{kin}_{exp}} \, \cdot
\, (4.60 \pm 0.56) \quad ,
\ee
which implies a theoretical error of about 12 \%.

To illustrate that our proposal has the potential to
lead to a more precise
determination of $V_{ub}$, it is instructive to briefly review the
present situation. The traditional
inclusive lepton endpoint spectrum analysis done at
CLEO \cite{cleoalt} and at ARGUS \cite{argusvub}, leads to
the result
$|V_{ub}|/|V_{cb}|=0.08 \pm 0.03$,
where the error is dominated by theory \cite{ballbraundosch}.
A recent new input to this quantity is provided by the measurements
of the exclusive semileptonic decays
$ B \to (\pi,\rho,\omega) \, \ell \nu$ \cite{CLEOexclalt,CLEOexclneu}.
The value for $|V_{ub}|$
quoted in the most recent analysis \cite{CLEOexclneu}
is $|V_{ub}|=(3.3 \pm 0.2^{+0.3}_{-0.4} \pm 0.7) \times 10^{-3}$, where
the errors are statistical, systematic and estimated model dependence,
after excluding models which are unable to predict the correct
$\pi/\rho$ ratio.

\section{Summary and Discussion}
In this paper we have proposed to measure the hadron energy spectrum
from inclusive semileptonic $B$ decays
in order to extract the CKM matrix element
$V_{ub}$ with improved precision.
The main advantage of our proposal is that
the energy window, in which
the $b \to c \ell^- {\overline \nu}$ transition is kinematically absent
$(E_{had} < m_D)$, is much wider than the
corresponing window
available in the traditional
lepton energy endpoint spectrum analysis.
Even after imposing a relatively high lower--cut at 
$E_{had}=1$ GeV (in order to avoid
the region of phase space where the range in the invariant
hadronic mass is too narrow to invoke quark-hadron duality),
a much larger fraction of the  $b \to u \ell^- {\overline \nu}$ 
events is captured in the remaining window $1 \,
\mbox{GeV} \le E_{had} \le m_D$
than in the lepton spectrum endpoint analysis.

After calculating the hadron energy spectrum in QCD improved
perturbation theory, we have implemented the
bound-state effects using both
the ACCMM  and the HQET approaches; the two methods  gave
essentially the same result, in particular
 in the relevant kinematical window.
Qualitatively, the dominant bound state  effect
is a uniform shift of the parton level spectrum.
We have pointed out that
the theoretical error in the ratio $|V_{ub}|/|V_{cb}|$, 
which is dominated
by the present uncertainties in
the $m_b$ quark mass, is about a factor of 2 to 3 smaller than the one
in the
lepton endpoint spectrum analysis and about a factor of 2
smaller than the model uncertainties in the
branching fractions of the exclusive decays $B \to
(\pi, \rho, \omega) \, \ell \nu$.
Therefore, for both statistical and theoretical reasons, the 
extraction of $|V_{ub}|$ from the hadron energy spectrum seems 
very attractive.

On the experimental side, our
proposal is most suited for a symmetric $B$-factory
running at the $\Upsilon(4S)$ resonance, which is
currently available at the
CLEO experiment.
Tagging the events from the
$\Upsilon(4S)$ decay, in which one $B$-meson is decaying
semileptonically and the other one nonleptonically,
the energy of the final state hadrons stemming from the semileptonically
decaying $B$-meson
is easily obtained by adding up the energies
of all the hadrons in the final state
and then subtracting $m_{\Upsilon}/2$.
On the other hand, in asymmetric $B$ factories, the hadron
energy spectrum is harder to measure because one first
has to reconstruct the
corresponding distribution in the rest frame of $\Upsilon (4S)$;
to perform the corresponding boost,
one has to measure precisely
both the energy and the momentum
of each final state hadron, which requires accurate particle
identification.

The spectrum of the invariant hadronic mass ($d\G/dm_{X}$)
in inclusive semileptonic decays
$B \to X_{c,u} \ell \nu$
is another source from
which one may try to extract $V_{ub}$.
Requiring $m_X$ to be below $m_D$, the
process $B \to X_c \ell \nu$ can be totally supressed.
Consequently, the integral of the hadron invariant mass distribution
below $m_D$ is another observable proportional to $|V_{ub}|^2$.
As the invariant mass is integrated over a large range,
 which covers
about 95 \%
of all the $b \to u \ell^- \overline{\nu}$ events \cite{Dai},
this observable is also theoretically viable.
However, we believe that the invariant mass spectrum is more 
difficult to measure than the hadron energy spectrum proposed in 
this paper, because first,
one has to measure the four momenta of all the final state
hadrons from the $\Upsilon(4S)$ decay and second,
one has to find a subset of final state hadrons with an invariant
mass of $m_B$, corresponing to the $B$-meson which decays
hadronically. Only then the invariant mass of the semileptonically
decaying $B$-meson can be determined.

Finally, we point out a potentially interesting
possibility of a  direct measurement of $\alpha_s (m_b)$ 
or ${\overline \Lambda}$. We have shown that, once the
real gluon bremsstrahlung correction is taken into account, 
the kinematic maximum of the hadron energy shifts from the 
tree--level (and virtual gluon correction) endpoint $(m_b^2 + m_q^2)/2m_b$ 
to $m_b$.  For the $b \rightarrow u$ case, 
Fig. 1 shows that the bremsstrahlung 
tail extends approximately 400 MeV beyond $m_b/2$. We also have shown that 
the dominant bound-state effect is to shift the spectrum by 
${\overline \Lambda}$ for both ACCMM and HQET approaches. Therefore, once 
${\overline \Lambda}$ is known accurately, then $\alpha_s (m_b)$ can be 
extracted directly from the bremsstrahlung tail spectrum. 
Alternatively, given an accurate value of $\alpha_s (m_b)$,
the non-perturbative parameter $\overline \Lambda$ can be 
accurately measured.
Since the bremsstrahlung spectrum of $b \rightarrow c$ extends further out
than that of $b \rightarrow u$, the above proposal should be most suited 
for the $b \to c$ transition.
In this case, one has to take into account the finite $c$-quark mass
dependence of the perturbative QCD corrections and bound-state effects.

\vskip0.5cm
We thank A. Ali, S.J. Brodsky, B. Grinstein, S.-K. Kim, 
J.H. K\"uhn, Y.-J. Kwon, M. Lu, 
P. Minkowski, 
R. Poling, E. Thorndike, and R. Wang for helpful discussions.

\end{document}